\documentclass[preprint,12pt]{elsarticle}

\usepackage{amssymb}

\journal{Physics Letter B}

\begin{document}

\begin{frontmatter}



\title{Constraints on nucleon effective mass splitting with Heavy Ion Collisions}


\author[1]{Yingxun Zhang\corref{cor1}}
\ead{zhyx@ciae.ac.cn}
\cortext[cor1]{Corresponding author}
\author[2] {M.B.Tsang} 
\author[1] {Zhuxia Li} 
\author[3] {Hang Liu}

\address[1] {China Institute of Atomic Energy, P.O. Box 275 (10), Beijing 102413, P.R. China}
\address[2] {National Superconducting Cyclotron Laboratory, Michigan State University, East Lansing, MI 48824, USA}
\address[3] {Texas Advanced Computing Center, University of Texas, Austin, Texas 78758, USA}

\begin{abstract}
A new version of the improved quantum molecular dynamics model has been developed to include standard Skyrme interactions. Four commonly used Skyrme parameter sets, SLy4, SkI2, SkM* and Gs are adopted in the transport model code to calculate the isospin diffusion observables as well as single and double ratios of transverse emitted nucleons. While isospin diffusion observables are sensitive to the symmetry energy term, they are not very sensitive to the nucleon effective mass splitting parameters in the interactions. Our calculations show that the high energy neutrons and protons and their ratios from reactions at different incident energies provide a robust observable to study the momentum dependence of the nucleon effective mass splitting. However the sensitivity of effective mass splitting effect on the n/p yield ratios decreases with increasing beam energy, even though high energy proton and neutron are produced more abundantly at high beam energy. Our calculations show that the optimum incident energy to study nucleon effective masses is between 100-200 MeV per nucleon.
\end{abstract}

\begin{keyword}
nucleon effective mass splitting, symmetry energy, heavy ion collisions


\end{keyword}

\end{frontmatter}


For nuclear matter with unequal number of neutrons and protons, the energy of the system is reduced by approximately the square of the differences in the proton and neutron densities divided by the total density. Such reduction in energy is called the nuclear symmetry energy which also appears in the nuclear binding energy of the liquid drop model. Thus the symmetry energy is of fundamental importance in our understanding of nature's asymmetric objects including neutron stars as well as heavy nuclei with very different number of neutrons and protons. Theoretical predictions on the symmetry energy have large uncertainties\cite{Brown00,BALi08}. This stimulates a lot of efforts in the nuclear physics communities to provide experimental constraints on the density dependence of symmetry energy. Observables used to constrain the symmetry energy range from isospin diffusions, yield ratios of emitted nucleons, isoscaling and flow of light charged particles in heavy ion collisions (HIC) \cite{BALi08, Tsang01,Tsang04, Famiano06, Ono03, Tsang09, Zhang08, Zhang12, Coupl11, SKumar11, LWChen05, BALi97, TXLiu07,Sun10,Kohley10}, to experiments that measure nuclei properties such as neutron skin\cite{Warda10, LWChen10, Gaidarov12}, Pygmy Dipole Resonance \cite{Carbonne,Wieland,Piekar11}, masses of Isobaric Analog States \cite{Danie09}, and nuclei masses\cite{Moller12,MLiu10}. Only recently, a consistent picture on the symmetry energy at saturation density, $S_0$, and its slope, $L$ has been obtained\cite{Tsang09,Tsang12}. The slope $L=3\rho_0 dS(\rho)/d\rho |_{\rho=\rho_0}$ where $S(\rho)$ is the density dependence of the symmetry energy is related to the pressure of pure neutron
matter at saturation density. Initial results from astrophysical measurements seem to favor much lower L values\cite{Stein12} than other experimental constraints. However, extraction of the radii from neutron stars is not settled and the latest results are more consistent with the heavy ion collision results\cite{Steiner}.

Another source of uncertainties in the theoretical description of the symmetry energy comes from the momentum dependence of the symmetry potential. Inside a strongly interacting medium as in the description of nuclear matter, the momentum dependence of symmetry potential may lead to different values of the neutron/proton effective masses. The knowledge of the difference,  also known as the nucleon effective mass splitting, is important for understanding not only level density of single particles, isovector Giant Dipole Resonance in nuclear structures and spectra of emitted particles in nuclear reactions but also many critical issues in astrophysics, such as heat capacity of matter, the neutron and proton chemical potential and their fraction \cite{Erler10,Lesin05,BALi08,Rizzo05,BALi04,BALi04m,BALi04npa,ZQFeng12,Bethe90,Farin01}.
There have been some efforts on constraining the nucleon effective mass splitting by analyzing the symmetry potenial $U_{sym} (\rho_0,E)$ using nuclei optical potential data\cite{BALi08,BALi04m,CXu10}. In that study, $m_n^*>m_p^*$ is found to be valid around normal density and Fermi momentum. However, other theoretical predictions including Relativistic Hartree Fock calculations\cite{WHLong08} suggest that at higher energy, $m^*_n>m^*_p$ change to $m^*_n<m^*_p$. Similar behaviors have also been found in Skyrme/Gogny-Hartree-Fock predictions on $U_{sym}(\rho,E)$\cite{RChen12}.

At high incident energy, HIC can create ``excited nuclear matter'' above the normal density and nucleon momentum distribution is far from the Fermi momentum for violent nucleon-nucleon collisions. To explore the effective mass splitting issues, we calculate the yield ratios of neutrons and protons emitted from HIC over a range of incident energies. In this paper, we explain how to constrain both the symmetry energy and nucleon effective mass from heavy ion collisions using transport model. Specifically, we incorporate Skyrme effective nucleon-nucleon interaction (or energy density functional) that has been used to describe nuclear structure properties \cite{Vauth} in the Improved Quantum Molecular Dynamics  (ImQMD) code. The constraints on the symmetry energy obtained with the new code are consistent with previous studies. Furthermore, our study suggests that nucleon yields from heavy ion collisions are sensitive to nucleon mass splitting at beam energy as low as 50 MeV per nucleon even though the optimum energy to study this effect is found to be around 100-200 MeV per nucleon, within the realms of current and future rare isotope beam facilities.


In most transport models, the nucleon energy density can be written as the local term and momentum dependent interaction term (MDI term) as
\begin{equation}
u=u_{loc}+u_{md}
\end{equation}
However, $u_{loc}$ in most transport codes adopts the Skyrme like energy density, $\frac{\alpha}{2}\frac{\rho^2}{\rho_0}+\frac{\beta}{\eta+1}\frac{\rho^{\eta+1}}{\rho_0^\eta}+S(\rho)\delta^2\rho$, while $u_{md}$ is often obtained by fitting the effective mass and energy dependence of the real part of the optical potential \cite{Aich, Hart, Bertch}. Most of these interactions used in transport models successfully describe the HIC observables, but few of them were used to study the nuclear structure. Exceptions are inclusion of the Gogny finite range interactions in the Asymmetrized Molecular Dynamic (AMD) and the IBUU04 codes \cite{Ono03,LWChen05}.
In this paper, we modified the nucleon potential part of the Improved Quantum Molecular Dynamics (ImQMD05) code\cite{Zhang05} to that derived from the real Skyrme energy potential energy density (without spin-orbit term) by including the energy density of isospin dependent Skyrme-like MDI as
\begin{eqnarray}
u_{md} &=& u_{md}(\rho\tau)+u_{md}(\rho_n\tau_n)+u_{md}(\rho_p\tau_p)\\\nonumber
&=& C_0 \int d^3p d^3p' f(\vec r,\vec p)f(\vec r,\vec p')(\vec p-\vec p')^2 +\\\nonumber
& &D_0 \int d^3p d^3p' [f_n(\vec r,\vec p)f_n(\vec r,\vec p')(\vec p-\vec p')^2\\\nonumber
&  & +f_p(\vec r,\vec p)f_p(\vec r,\vec p')(\vec p-\vec p')^2]
\end{eqnarray}
where $f(\vec r,\vec p)$ is the nucleon phase space density, and $f(\vec r,\vec p)= \sum_i \frac{1}{(\pi\hbar)^3}exp[-(\vec r-\vec r_i)^2/2\sigma_r^2-(\vec p-\vec p_i)^2/2\sigma_p^2]$ in QMD approaches. The coefficients $C_0$ and $D_0$ can be determined with following relationship,
\begin{eqnarray}
C_0 &=&\frac{1}{16\hbar^2}[t_1(2+x_1)+t_2(2+x_2)]\\
D_0 &=&\frac{1}{16\hbar^2}[t_2(2x_2+1)-t_1(2x_1+1)]
\end{eqnarray}
For nuclear matter at zero temperature, $f_q=\frac{2}{(2\pi\hbar)^3 }\theta(p-p^q_F)$, $q=n, p$.  Eq.(2) can be calculated analytically, and it is equal to $A_0\rho\tau+B_0 (\rho_n\tau_n+\rho_p\tau_p)$ (as same as $\rho^{8/3}$ and $\rho^{8/3}\delta^2$ term in Eq.(2) in Ref.\cite{Zhang05}), where $\tau=\tau_n+\tau_p$, $\tau_q=\frac{3}{5} k_{q,F}^2\rho_q$. $A_0=1/8[t_1 (2+x_1 )+t_2 (2+x_2)]$, $B_0=1/8[t_2 (2x_2+1)-t_1 (2x_1+1)]$.
The local part of energy density is defined as:
\begin{eqnarray}
u_{\rho}&=&\frac{\alpha}{2}\frac{\rho^2}{\rho_0}+
\frac{\beta}{\eta+1}\frac{\rho^{\eta+1}}{\rho_0^{\eta}}+
\frac{g_{sur}}{2\rho_0}(\nabla\rho)^2\nonumber\\
&  &+\frac{g_{sur,iso}}{\rho_0}[\nabla(\rho_n-\rho_p)]^2\nonumber\\
&  &+A_{sym}\rho^2\delta^2
+B_{sym}\rho^{\eta+1}\delta^2
\end{eqnarray}
here, $\delta=(\rho_n-\rho_p)/(\rho_n+\rho_p)$ is the isospin asymmetry, $\rho_n$ and $\rho_p$ are the neutron and proton densities, respectively. The coefficients of $\alpha$, $\beta$, $g_{sur}$, $g_{sur,iso}$, $A_{sym}$ and $B_{sym}$ can be obtained by the standard Skyrme interaction parameters as in previous work \cite{Zhang05}.
These calculations use isospin-dependent in-medium nucleon nucleon scattering cross sections in the collision term and Pauli blocking effects as described in \cite{Zhang05}. This new version of Quantum Molecular Dynamics code (ImQMD-Sky) retains the many-body correlations incorporated in the original QMD approaches\cite{Aich, Wang02}.

In the following studies, we choose four Skyrme interaction parameter sets, SLy4, SkI2, SkM* and Gs \cite{Chabn97, Reinhard95, Bartel, Friedrich} which have similar incompressibility ($K_0$), symmetry energy coefficient ($S_0$) and isoscalar effective mass ($m^*$), i.e., $K_0=230\pm20 MeV$,  $S_0=32\pm2MeV$ and $m^*/m=0.7\pm0.1$. The SLy4 and SkI2\cite{Chabn97, Reinhard95} have similar nucleon effective mass splitting, with $m_n^*<m_p^*$, but very different slopes of symmetry energy $L$ values, 46MeV for SLy4, and 104MeV for SkI2. The other two Skyrme interaction parameter sets with $m_n^*>m_p^*$ also have different $L$ values, 46MeV for SkM*\cite{Bartel}, and 93MeV for Gs\cite{Friedrich}. The saturation properties of nuclear matter for these four Skymre interactions are listed in Table.I. By analyzing the results from calculations that use these interactions, we hope to disentangle the sensitivities of the isospin observables on the density dependence of symmetry energy and neutron proton effective mass splitting.

\begin{table}[b]
\caption{\label{tab:table1}%
Corresponding saturation properties of nuclear matter in, SLy4, SkI2, SkM*, and Gs Skyrme parameters. All entries are in MeV, except for $\rho_0$ in $fm^{-3}$ and the dimensionless effective mass ratios for nucleon, neutron and proton. The effective mass for neutron and proton are obtained for isospin asymmetric nuclear matter with $\delta=0.2$.}
\begin{tabular}{lccccccccc}
\hline
\hline
\textrm{Para.}&
\textrm{$\rho_0$}&
\textrm{$E_0$}& \textrm{$K_0$} & \textrm{$S_0$} & \textrm{$L$} & \textrm{$K_{sym}$} &
\textrm{$m^*/m$} &$m^*_n/m$ & $m^*_p/m$\\
\hline
 SLy4 & 0.160 & -15.97 & 230 &  32 & 46 & -120& 0.69& 0.68 &0.71 \\
 SkI2 & 0.158 & -15.78 & 241 &  33 & 104 & 71 & 0.68& 0.66 &0.71 \\
 SkM* & 0.160 & -15.77 & 217 &  30 & 46 & -156 & 0.79 & 0.82 &0.76 \\
 Gs   & 0.158 & -15.59 & 237 &  31 & 93 & 14 & 0.78 & 0.81 &0.76 \\
\hline
\end{tabular}
\end{table}

The left panel of Figure 1 shows the density dependence of symmetry energy for cold nuclear matter, $ S(\rho)=\frac{1}{3}\frac{\hbar^2}{2m}\rho_0^{2/3}(\frac{3\pi^2}{2}\frac{\rho}{\rho_0})^{2/3}+A_{sym}\rho
+ B_{sym}\rho^\eta+C_{sym}\rho^{5/3}$. The last term in $S(\rho)$ is derived from the $u_{md}$ in Eq.(2), and $C_{sym}=-\frac{1}{24}(\frac{3\pi^2}{2})^{2/3}\Theta_{sym}$, $\Theta_{sym}=3t_1 x_1-t_2 (4+5x_2)$. Smaller $L$ values yield higher symmetry energy at subsaturation densities while the opposite is true at the suprasaturation density regions. The right panels of Figure 1 show the Lane potentials $U_{sym}= \frac{U_n-U_p}{2\delta}=2A_{sym}\rho+2B_{sym}\rho^\eta+2D_0 m\rho E_k$ for cold nuclear matter at $0.5\rho_0$ (top panel) and at $\rho_0$ (bottom panel) as a function of nucleon kinetic energy. The Lane potential gives an accurate estimate for the difference of the force between neutron and proton experienced in asymmetric nuclear matter, and directly influences the neutron proton yield ratios, Y(n)/Y(p). The larger the Lane potential, the larger is the Y(n)/Y(p) ratio. A crossing of two Lane potentials for SLy4 (SkI2) and SkM*(Gs) appears at $\sim30MeV$ at $\rho=0.5\rho_0$ and $\sim70MeV$ at $\rho=\rho_0$. The cross-over is especially clear at the higher density region. Thus, we can expect the Y(n)/Y(p) ratios to have a cross-over point at high enough density regions.
\begin{figure}[htbp]
\centering
\includegraphics[angle=270,scale=0.5]{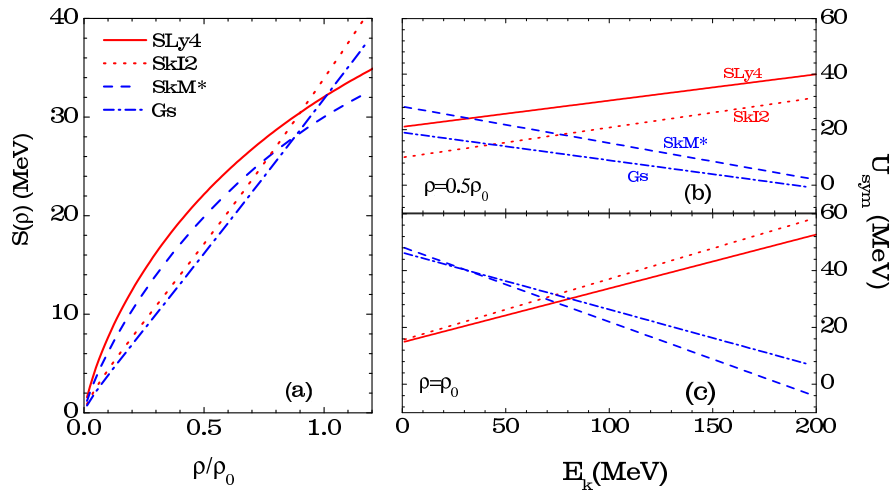}
\setlength{\abovecaptionskip}{50pt}
\caption{\label{ref-fig1}(Color online) (a) Density dependence of symmetry energy, (b) Energy dependence of the Lane potential at $\rho=0.5\rho_0$ and (c) $\rho=\rho_0$, for the parameters SLy4 (solid line), SkI2 (dot line), SkM*(dash line) and Gs (dot dash line).}
\setlength{\belowcaptionskip}{0pt}
\end{figure}

In previous studies \cite{Tsang04,Tsang09, Zhang12, LWChen05, Coupl11}, particles emitted from the neck region were used to measure the isospin diffusion. The isospin transport ratio $R_i$ was introduced to quantify the isospin diffusion effects \cite{Tsang04,Shi03},
\begin{equation}
R_i=\frac{2X-X_{aa}-X_{bb}}{X_{aa}-X_{bb}}
\end{equation}
where $X$ is an isospin observable and the subscripts $a$ and $b$ represent the neutron-rich and neutron-poor nuclei. In this work, we use $a$ and $b$ to denote the projectile (first index) and target (second index) combination, where $a = ^{124}Sn$ and $b = ^{112}Sn$. We obtain the value of $R_i$ by three reaction systems, $a + a$, $b + b$, and $a + b$ (or $b + a$). In the absence of isospin diffusion, the ratios are $R_i(X_{ab})=R_i (X_{aa})=1$ and  $R_i (X_{ba} )=R_i (X_{bb} )=-1$. If isospin equilibrium is achieved, then $R_i (X_{ab} )=R_i (X_{ba} )\approx0$. Until now, the best set of data from heavy ion collisions consists of the isospin diffusion transport ratios obtained from isoscaling parameters $X=\alpha$ and from the rapidity dependence of the yield ratios of $A=7$ mirror nuclei, $X_7=ln(Y(^7 Li)/Y(^7 Be))$ \cite{TXLiu07,Sun10} from $^{124,112}Sn+^{124,112}Sn$ collisions at beam energy of 50 MeV per nucleon\cite{Tsang09,LWChen05}. Even though there are data from isoscaling and fragment flow, they are not as thoroughly studied with transport models and the extracted constraints have been used mainly for consistency checks.

In this paper, we simulated the collisions of $^{124}Sn+^{124}Sn$, $^{124}Sn+^{112}Sn$, $^{112}Sn+^{124}Sn$, and $^{112}Sn+^{112}Sn$ reactions at beam energy of 50 AMeV using the ImQMD-Sky code. 64,000 events are performed for each reaction at each impact parameter. Previous theoretical studies\cite{Zhang12,Coupl11} and recent experimental studies\cite{Coupland} suggest that there is no strong dependence on the impact parameter. In the left panel of Figure 2, we plot the isospin transport ratios obtained with SLy4, SkI2, SkM* and Gs interactions at b=6fm. As in previous studies \cite{Tsang09, Zhang12}, we analyze the amount of isospin diffusion by constructing a tracer, $X=\delta$, from the isospin asymmetry of emitting source which includes all emitted nucleons (N) and fragments (frag) with the velocity cut ($v_z^{N,frag}>0.5v_{beam}^{c.m.}$  ). The shaded region is experimental data obtained by constructing the isospin transport ratio using isoscaling parameter X=$\alpha$, near the projectile rapidity regions\cite{Tsang04}. Our results show that the $R_i$ values for SLy4 (solid circle) and SkM* (solid squares), both with $L=46 MeV$, lie within the experimental uncertainties while the $R_i$ values for SkI2 (open circle, $L=104 MeV$) and Gs (open square, $L=93 MeV$) are above the data range. Even though the isospin diffusion process is accelerated at subsaturation densities with the stronger Lane potential, the overall effect of mass splitting on isospin diffusion is small. This conclusion is similar with previous results even from the IBUU and SMF models\cite{LWChen05,Rizzo08}. Since the isospin diffusion process is strongly related to the difference of isospin concentration and the strong repulsive momentum-dependent isoscalar potential reduces the effect of isovector potential on the reaction dynamics, thus, there is no clear pattern that $R_i$ values decrease significantly with the strength of Lane potentials.

\begin{figure}[htbp]
\centering
\includegraphics[angle=270,scale=0.5]{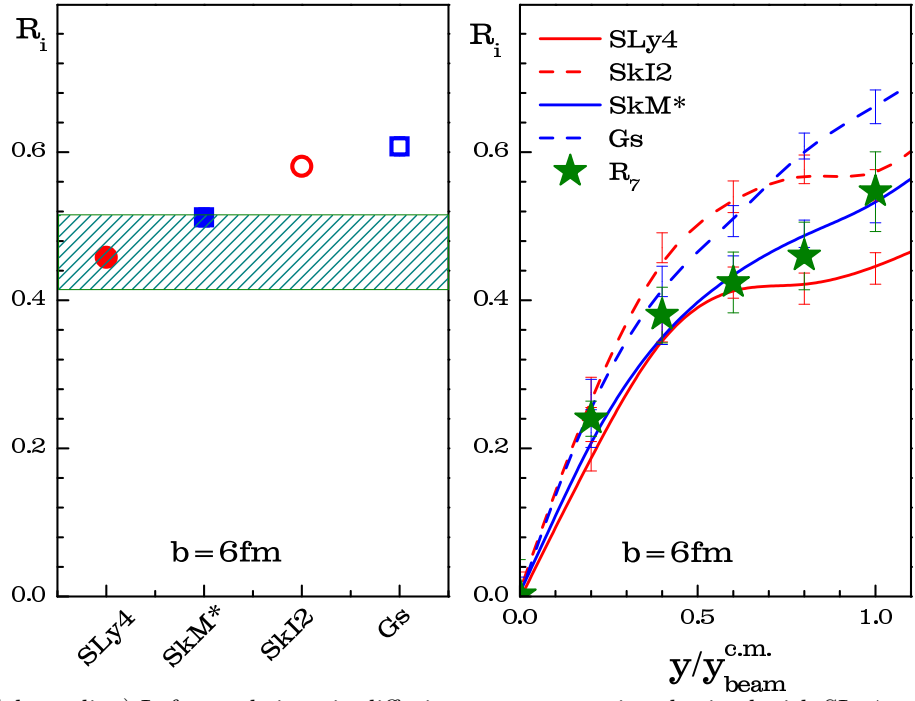}
\setlength{\abovecaptionskip}{50pt}
\caption{\label{ref-fig2}(Color online) Left panel: isospin diffusion transport ratios obtained with SLy4, SkI2, SkM* and Gs. The shaded region corresponds to the data from \cite{Tsang04}. Right panel: isospin transport ratios as a function of rapidity for SLy4, SkI2, SkM* and Gs. The star symbols are data from\cite{TXLiu07}.}
\setlength{\belowcaptionskip}{0pt}
\end{figure}

We also compare results of the calculations to $R_i$ as a function of the scaled rapidity $y/y_{beam}^{c.m.}$ as shown in the right panel of Figure 2. The star symbols in the right panel are experimental data, $R_i(X_7)=R_7$, obtained in Ref. \cite{TXLiu07}. This transport ratio was generated using the isospin tracer X = ln[Y ($^7$Li)/Y($^7$Be)], where Y ($^7$Li)/Y($^7$Be) is the yield ratio of the mirror nuclei, $^7$Li and $^7$Be \cite{TXLiu07}. For comparison, the ImQMD-Sky calculations of $R_i$ are plotted as lines for b=6fm. The interactions with smaller $L$ values, SLy4 and SkM* (solid lines) agree with the data better especially in the high rapidity region. However, $\chi^2$ analysis suggests that the quality of fit with isospin diffusion data is not good enough to draw definite conclusions about mass splitting effect with confidence. We need a more sensitive and reliable observable to extract quantitative information about the nucleon effective mass splitting.


Since the contributions from MDI part of symmetry potential play more important roles on the emissions of neutrons and protons yield ratios when the relative momentum of nucleons increase. The yield ratios of Y(n)/Y(p) should be sensitive to the effective mass splitting, and it had been found in \cite{Rizzo05}. Important information about the strength of symmetry potential at high kinetic energy(or the sign of effective mass splitting at high momentum) can be obtained from the pre-equilibrium transverse emission of nucleons at earlier stage of reactions. In Figure 3, we plot the $Y(n)/Y(p)$ ratios as a function of nucleon center of mass energy, $E_{c.m.}$, for $^{112}Sn+^{112}Sn$ (left panel) and $^{124}Sn+^{124}Sn$ (middle panel) at b=2fm with angular gate $70^\circ<\theta_{c.m.}<110^\circ$. The lines connecting the circles correspond to $m_n^*<m_p^*$ case, and the lines connecting the squares correspond to $m_n^*>m_p^*$ case. Not surprisingly, the $Y(n)/Y(p)$ ratios are larger for the neutron rich system, $^{124}Sn+^{124}Sn$, in the middle panel. Consistent with Ref. \cite{BALi04npa, Rizzo05, BALi04,ZQFeng12}, the differences in the $Y(n)/Y(p)$ ratios between the $m_n^*<m_p^*$ (circles) and $m_n^*>m_p^*$ (squares) increase with nucleons kinetic energy. At high nucleon energies, the stronger Lane potentials with $m_n^*<m_p^*$ enhance neutron emissions, leading to flatter Y(n)/Y(p) dependence on the nucleon kinetic energy.

\begin{figure}[htbp]
\centering
\includegraphics[angle=270,scale=0.5]{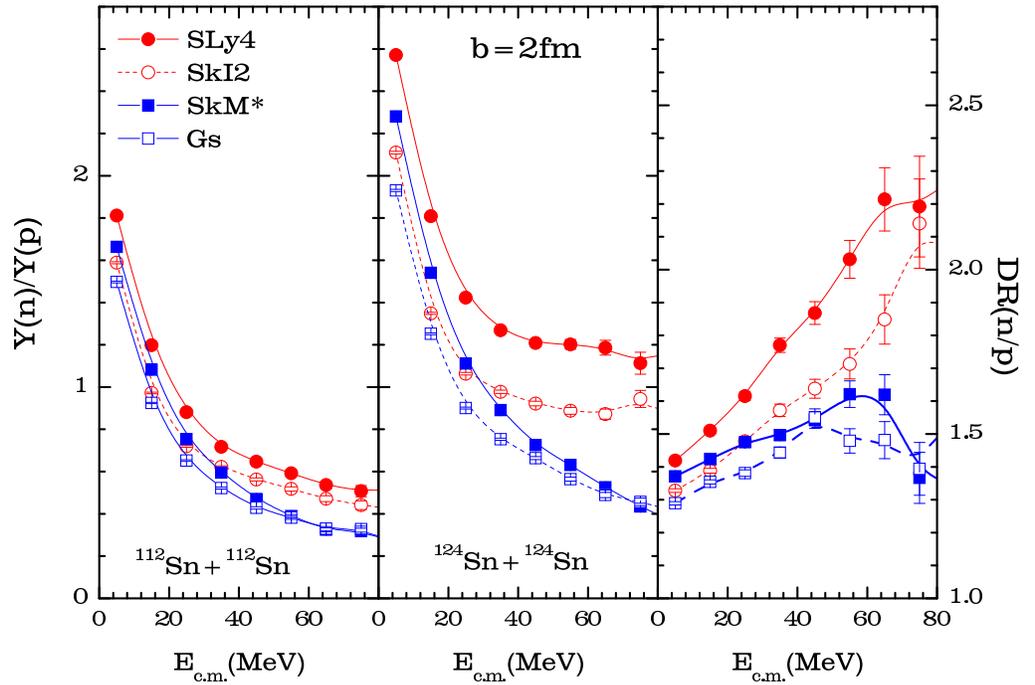}
\setlength{\abovecaptionskip}{50pt}
\caption{\label{ref-fig3} (Color online) Left panel: $Y(n)/Y(p)$ as a function of kinetic energy for $^{112}Sn+^{112}Sn$ at b=2fm with angular cuts $70^\circ<\theta_{c.m.}<110^\circ$; Middle panel is the $Y(n)/Y(p)$ for $^{124}Sn+^{124}Sn$. Right panel: DR(n/p) ratios as a function of kinetic energy. The calculated results are for SLy4 (solid circles), SkI2 (open circles), SkM* (solid squares) and Gs (open squares)}
\setlength{\belowcaptionskip}{0pt}
\end{figure}

The calculated results on double ratio DR(n/p)\cite{Double} are shown in the right panel of Figure 3. The calculations with SLy4($L=46 MeV,m_n^*<m_p^*$) are consistent with the double ratios data from Ref. \cite{Famiano06}, especially at high kinetic energy region. However these data points have very large uncertainties. Furthermore, recent remeasurements of the data suggest different trends and different values of DR(n/p) \cite{Coupland}. Until the experimental results are finalized, we decide not to make comparisons between data and calculations.

Since the effect of mass splitting should be larger at higher densities and the sign of splitting may change at high momentum, we explore $Y(n)/Y(p)$ ratios at $E_{beam}$=100, 150, 200 and 300 AMeV with b=2 fm for $^{124}Sn+^{124}Sn$ and $^{112}Sn+^{112}Sn$. For clearer presentation, we restrict our calculations to the SLy4 and SkM* interactions which have opposite mass splitting with the same $L=46MeV$ values. (SkI2 and Gs show similar sensitivities to the effective mass-splitting as SLy4 and SkM* respectively as demonstrated in the right panel of Figure 3.) In the top panel of Figure 4, we present $Y(n)/Y(p)$ ratios as a function of $E_{c.m.}$ for the neutron-rich system, $^{124}Sn+^{124}Sn$. Consistent with the lower right panel of Figure 1, a cross over of Y(n)/Y(p) is observed from low to high nucleon energy between SLy4 interactions with $m_n^*<m_p^*$ and SkM* with $m_n^*>m_p^*$ when E$_{beam}>200$AMeV. This behavior is similar with the results presented in \cite{Giord10}, but the values are different between the two codes. It may be caused by the different form of MDI used in the codes.

The corresponding double ratios $DR(n/p)$ are plotted as a function of nucleon center of mass energy in the bottom panel of Figure 4. $DR(n/p)$ are nearly the same for nucleons emitted at $E_{c.m.}<50 MeV$, independent of incident energy and mass splitting. For $E_{c.m.}>50 MeV$,  $DR(n/p)$ increase with nucleon kinetic energy for SLy4 where $m_n^*<m_p^*$ while the trend is nearly flat or decrease slightly for SkM* with $m_n^*>m_p^*$. The sensitivity of the DR(n/p) ratios decrease with incident energies due to increase in nucleon-nucleon scattering. The current calculations suggest that the mass splitting is best studied at incident energy less than 200 MeV.
\begin{figure}[htbp]
\centering
\includegraphics[angle=270,scale=0.5]{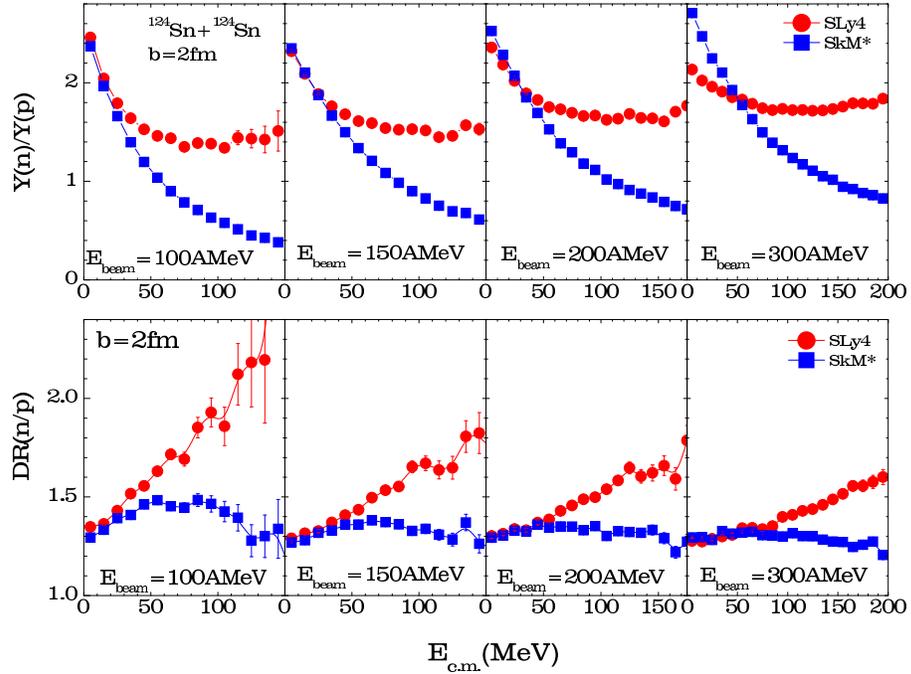}
\setlength{\abovecaptionskip}{50pt}
\caption{\label{ref-zweight-rapidity-fig5}(Top panels) Y(n)/Y(p) ratios as a function of kinetic energy for $^{124}Sn+^{124}Sn$ at b=2fm; (Bottom panels) DR(n/p) ratios as a function of kinetic energy. From left to right, the beam energies are 100AMeV, 150AMeV, 200AMeV, 300AMeV.}
\setlength{\belowcaptionskip}{0pt}
\end{figure}

In summary, we have developed a new version of the improved quantum molecular dynamics code, which can accommodate real Skyrme interaction parameters (without spin-orbit term), to describe isospin diffusion as well as single and double neutron-proton ratios. We find that the mass splitting affects isospin diffusion minimally, and the isospin diffusion data favor the Skyrme interaction with low L values. We also show that the high energy neutrons and protons and their ratios from heavy ion reactions at different incident energies, especially around 100-200 MeV per nucleon, provide a good observable to study the momentum dependence of the nucleon effective mass splitting. Finally new neutron and proton spectral data with much smaller uncertainties than previous data at different beam energies may allow one to determine the magnitude of the nucleon mass splitting and its momentum dependence.


\textbf{Acknowledgements}
The authors thank Prof. Bao-An Li for helpful comments.
This work has been supported by the Chinese National Science Foundation under Grants (11075215, 10875031, 11005022, 11005155, 11275052), the 973 Program of China No. 2013CB834404, the Nuclear Energy Development and Research program in China No.[2011]767 and the USA National Science Foundation Grants No. PHY-1102511.






\begin{thebibliography}{00}
\bibitem{Double} To minimize the effects of detector efficiencies, the yield ratio from an asymmetric
system $A=^{124}Sn+^{124}Sn$ is compared to the yields from a more symmetric system $B=^{112}Sn+^{112}Sn$ by constructing the double ratio, $DR(n/p)=Y(n)/Y(p)_{124}/Y(n)/Y(p)_{112}$.
\end{thebibliography}



\end{document}